\documentclass[twocolumn,twoside,slac_two]{revtex4}
\usepackage{graphicx}
\usepackage{fancyhdr}
\usepackage{xspace}
\newcommand{\dzero}{D\O\xspace}
\newcommand{\mtop}{\ensuremath{m_{\rm top}}\xspace}

\newcommand{\ljets}{\ensuremath{\ell}+\rm jets\xspace}
\newcommand{\ttbar}{\ensuremath{t\bar{t}}\xspace}
\newcommand{\ppbar}{\ensuremath{p\bar{p}}\xspace}

\begin{document}

\title{Top Mass Measurements at the Tevatron}


\author{M.H.L.S. Wang for the \dzero and CDF collaborations}
\affiliation{Fermi National Accelerator Laboratory, Batavia, IL 60510, U.S.A.}

\begin{abstract}
We present the latest measurements of the top quark mass from the Tevatron.  The
different top decay channels and measurement techniques used for these results
are also described.  The world average of the top quark mass based on some of
these new results combined with previous results is $\mtop=172.6\pm1.4$ GeV. 
\end{abstract}

\maketitle

\thispagestyle{fancy}


\section{Introduction}
The discovery of the top quark by the \dzero and CDF collaborations over a
decade ago was one of the crowning achievments of Fermilab's Tevatron
collider~\citep{topdiscovery}.  It was a testament to the tremendous success of
the standard model as a theory of particle interactions.  Over the years, much
effort has been devoted to the measurement and understanding of its properties. 
Of these, the single property that has perhaps received the most attention is
the mass.  Over 30 times larger than the $b$ quark mass, the sheer magnitude of
the top quark mass suggests a special role in the mechanism of electroweak
symmetry breaking.  The current interest in a precise measurement of the top
quark mass, together with that of the $W$ boson mass, is driven by the fact that
they provide powerful constraints on the mass of the Higgs boson through
radiative corrections, indicating the range in which to conduct
searches~\citep{lepewwg}.  In addition, it can also help constrain possible
extensions of the standard model.

We present, in this talk, the latest measurements from \dzero and CDF of the top
quark mass using different analysis techniques and top quark decay channels. 
These results are based on up to 2 fb$^{-1}$ of data collected per experiment at
Fermilab's Tevatron.

\section{Decay Channels}
The top quark mass measurements presented here are based on top quarks that are
produced as \ttbar pairs via the strong interaction.  In the Standard Model,
each top quark in the pair decays into a $W$ boson and a $b$ quark nearly 100\% of
the time resulting in a $W^{+}W^{-}b\overline{b}$ final state.  The subsequent
decays of the two $W$ bosons define the different top quark decay channels or
topologies.

The first of these decay channels is the dilepton channel in which both $W$
bosons decay ``leptonically" into a lepton and its associated neutrino.  This is
the cleanest channel with the lowest background levels dominated by Drell-Yan
processes with associated jets.  Additional sources include diboson production
with jets, and $W+\geq3$ jet and multijet production when one or 2 jets are
misidentified as leptons. Despite major drawbacks which include the lowest
branching fraction ($\sim5$\% for $e\nu$ and $\mu\nu$ modes) and the presence of
two undetected neutrinos, it is an important channel because the presence of
fewer jets and the sensitivity to different background sources provide a useful
cross check to measurements obtained from the other channels.

The second decay channel is the hadronic channel in which both $W$ bosons decay
``hadronically" into two quarks resulting in at least six jets in the event. 
The main advantages of this channel are the fact that it has the largest
branching fraction ($\sim46$\%) and that the jets evolving from all six quarks
in the final state can, in principle, be measured.  Furthermore, because the two
$W$ bosons decay hadronically, the well known mass of the $W$ boson can be used
to constrain the measured jet energies in order to reduce the uncertainty on the
jet energy scale which is a dominant source of systematic uncertainty.  The
major disadvantages of this channel are the large amount of background from
multijet events and the combinatoric problem when assigning jets to parton
identities.

The final decay channel is the lepton+jets (\ljets) channel where one $W$ boson decays
hadronically and the other leptonically.  Drawing from the strengths of the
other two channels, it combines a decent branching fraction of $\sim29$\% and a
manageable level of background from $W+$jets and multijet events.  It also
benefits from an \emph{in-situ} jet energy calibration due to the presence of a
hadronically decaying $W$ boson.  This channel has yielded the most precise
measurements of the top quark mass.

\section{A Challenging Measurement}
Top quark mass measurements are challenging because, except for the leptons from
the $W$ boson decay in two of the channels, the daughters of the top and $W$
boson are not identified directly.  What is measured in the detector are the
high transverse momentum jets evolving from the quarks and a large amount of
missing energy signifying the presence of neutrinos.  In general, one does not
know how to associate the jets with the quarks and must try all possible
combinations.  The use of $b$-tagging helps greatly in reducing the
combinatorics.  Although detached vertices are associated with the $b$-jets,
there are none associated with the short-lived top quark itself that can be used
to cleanly identify daughters in the same way they are used, for instance, in
mass measurements involving long lived particles like hyperons.  In the
following section, we discuss the different analysis techniques that make
precise measurements of the top quark mass possible despite such challenges.

\section{Analysis Techniques}

One of the most commonly used analysis techniques is the template method.  This
method makes use of a variable that is sensitive to the top quark mass.  An
obvious choice of this variable used by many analyses is the reconstructed top
quark mass.  Monte Carlo (MC) samples of \ttbar events generated with different
values of the top quark mass are then used to produce distributions of the
selected variable called templates.  Since each of these templates is associated
with a particular value of the true top quark mass used to generate the events,
one can extract the top quark mass from the data by comparing each of the
templates to the corresponding data distribution.  To do this, one can either
fit the templates directly to the data distribution or parametrize the templates
as probability density functions from which likelihoods can be constructed
that are used to extract the mass.

The analysis technique that has yielded the most precise top quark mass
measurements to date is the Matrix Element (ME) method which was pioneered by
\dzero in Run I of the Tevatron using the \ljets
channel~\citep{topmassd0nature}.  In this method, the probability to observe a
given event is calculated by taking into account the different physics processes
that could contribute to the observed event.  The total probability is simply
the properly normalized sum of the differential cross sections for each
contributing process calculated from the MEs characterizing the process which
gives the technique its name.  The probabilities are calculated for each event
as a function of the parameters one wishes to extract such as the top quark mass
and jet energy scale factor.  A joint likelihood is constructed from the
probability distribution of each event from which the best estimate of each
parameter and its uncertainty can be determined.  The power of this technique is
based on the maximal use of the available kinematic information in order to
fully specify an event.  While providing discrimination between different events
it also deals with the combinatoric problem within each event mentioned above by
properly weighting each jet permutation based on its probability.  Furthermore,
detector resolution effects are also accounted for with transfer functions which
give the probability density for a measured value of a given quantity as a
function of the true parton value. 

Both techniques described above rely on fully simulated MC \ttbar events for
calibrating the procedure in order to determine the true value of the top quark
mass from the extracted value.  The MC generators currently used for this
purpose are all based on leading order (LO) matrix elements with higher orders
simulated by parton showers leading to a concept of the top quark mass that is
not well defined.  To address this, a third technique extracts the top quark
mass by comparing the theoretically predicted and experimentally measured values
of the \ttbar production cross section.  Theory and measurement likelihoods which
are defined as a function of the top quark mass and \ttbar production cross
section are multiplied together to construct a joint likelihood. Integrating
over the \ttbar production cross section then gives a one dimensional likelihood
from which the top quark mass is extracted.

Finally a fourth technique addresses the fact that most  top quark mass
measurements are dominated by jet energy scale uncertainties.  Assuming top
quarks in \ppbar collisions are produced nearly at rest, it can be shown that the
Lorentz boost imparted to the $b$ quark is a function of the top quark mass. 
Because of this, the velocity of the $b$ quark and hence the $b$ hadron is
strongly correlated with the top quark mass. Therefore, the average momentum
of the $b$ hadron can be used to determine the top quark mass.  In practice,
this technique uses the highly correlated average transverse decay length of the
$b$ hadron instead of its average momentum.  The dependence of the decay length
on the top quark mass is parameterized with MC events and this parameterization
is used to extract the top quark mass from data.  This technique depends mainly
on tracking to determine the decay length precisely and is largely insensitive
to the jet energy scale uncertainties that dominate other techniques.

\section{Results from the Tevatron}

\begin{figure*}[t]
\centering
\includegraphics[width=60.mm]{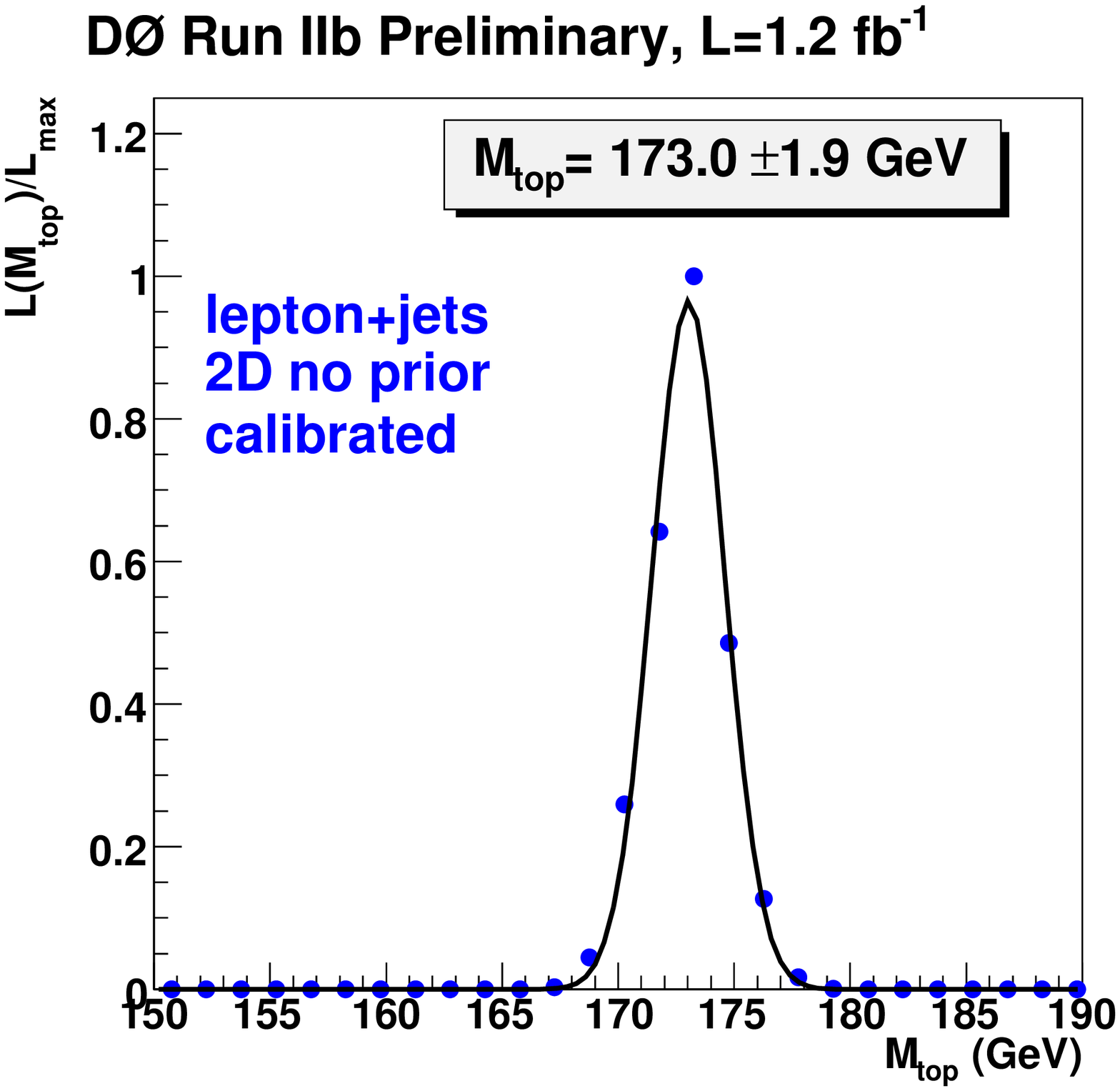}
\includegraphics[width=60.mm]{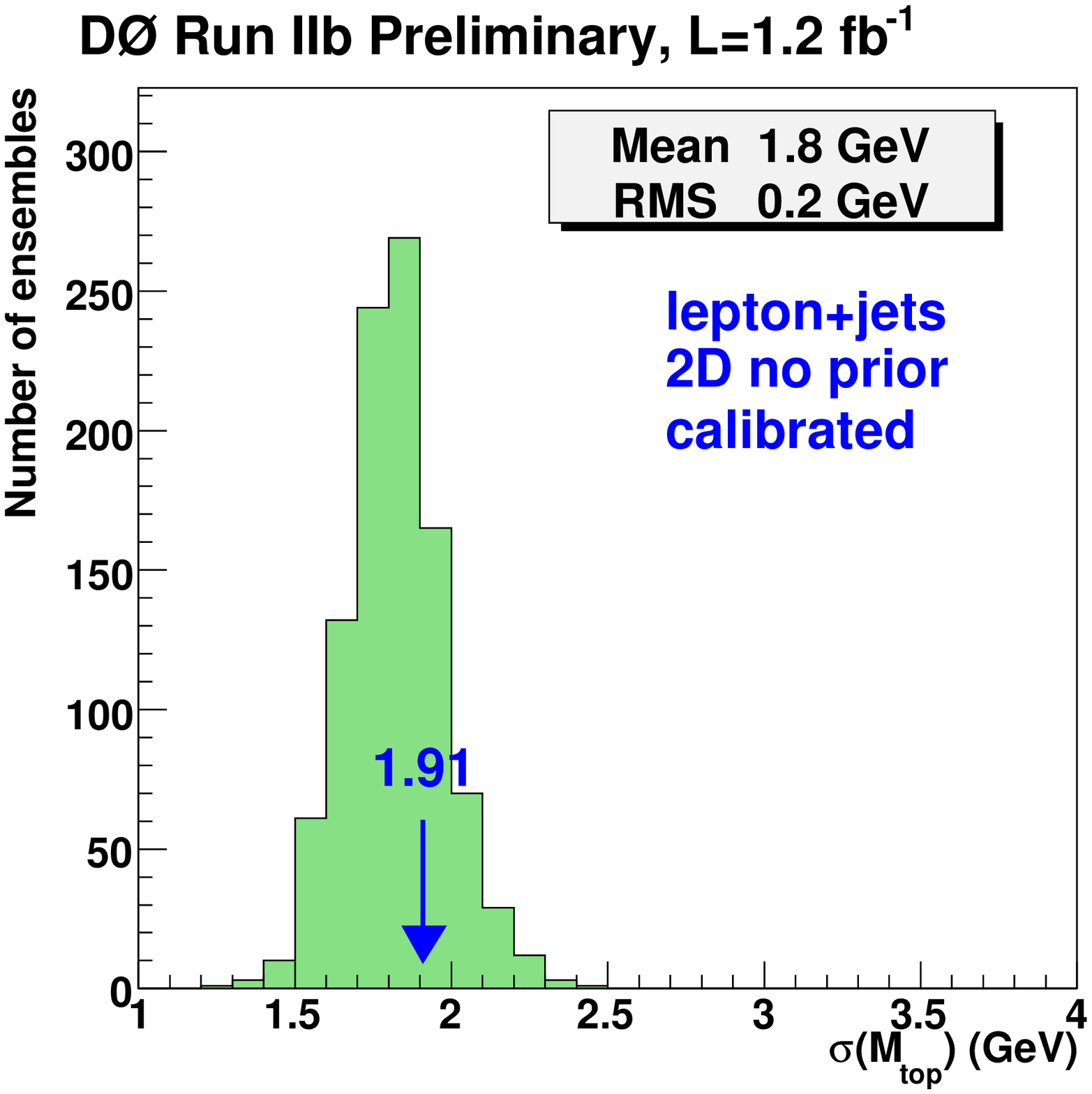}
\caption{\dzero result in the \ljets channel using the ME method. Likelihood as
a function of top quark mass for first 1 fb$^{-1}$ of data (left).  Expected
uncertainty distribution from MC ensemble tests (right).}
\label{fig_d0_meljets}
\end{figure*}
The first result is from \dzero in the \ljets channel~\citep{d0_meljets}.  It
uses the ME technique and is based on 2.1 fb$^{-1}$ of data.  The well known $W$
mass is used as constraint to perform an \emph{in-situ} jet energy calibration
through the introduction of a $JES$ parameter which is a global factor
multiplying the energies of all the jets.  The top quark mass (\mtop) is then
extracted by maximizing the joint likelihood simultaneously in terms of the top
quark mass and the $JES$ parameter.  Figure \ref{fig_d0_meljets} (left) shows
a one dimensional projection of the likelihood onto the top quark mass axis for
the second 1 fb$^{-1}$ of data yielding
$\mtop=173.0\pm1.9(\mbox{stat+JES})\pm1.0(\mbox{syst})$ GeV.  Figure
\ref{fig_d0_meljets} (right) shows the corresponding distribution of expected
uncertainties from MC ensemble tests with the arrow indicating the measurement
uncertainty.  The result for the first 1 fb$^{-1}$ of data is
$\mtop=170.5\pm2.5(\mbox{stat+JES})\pm1.4(\mbox{syst})$ GeV.  Combining both
results gives $\mtop=172.2\pm1.1(\mbox{stat})\pm1.6(\mbox{syst})$ GeV for the
entire data set.  The systematic uncertainty is dominated by the uncertainty on
the $b$ jet energy scale and the uncertainty in the modeling of extra jets from
radiation.

\begin{figure*}[t]
\centering
\includegraphics[width=80.mm]{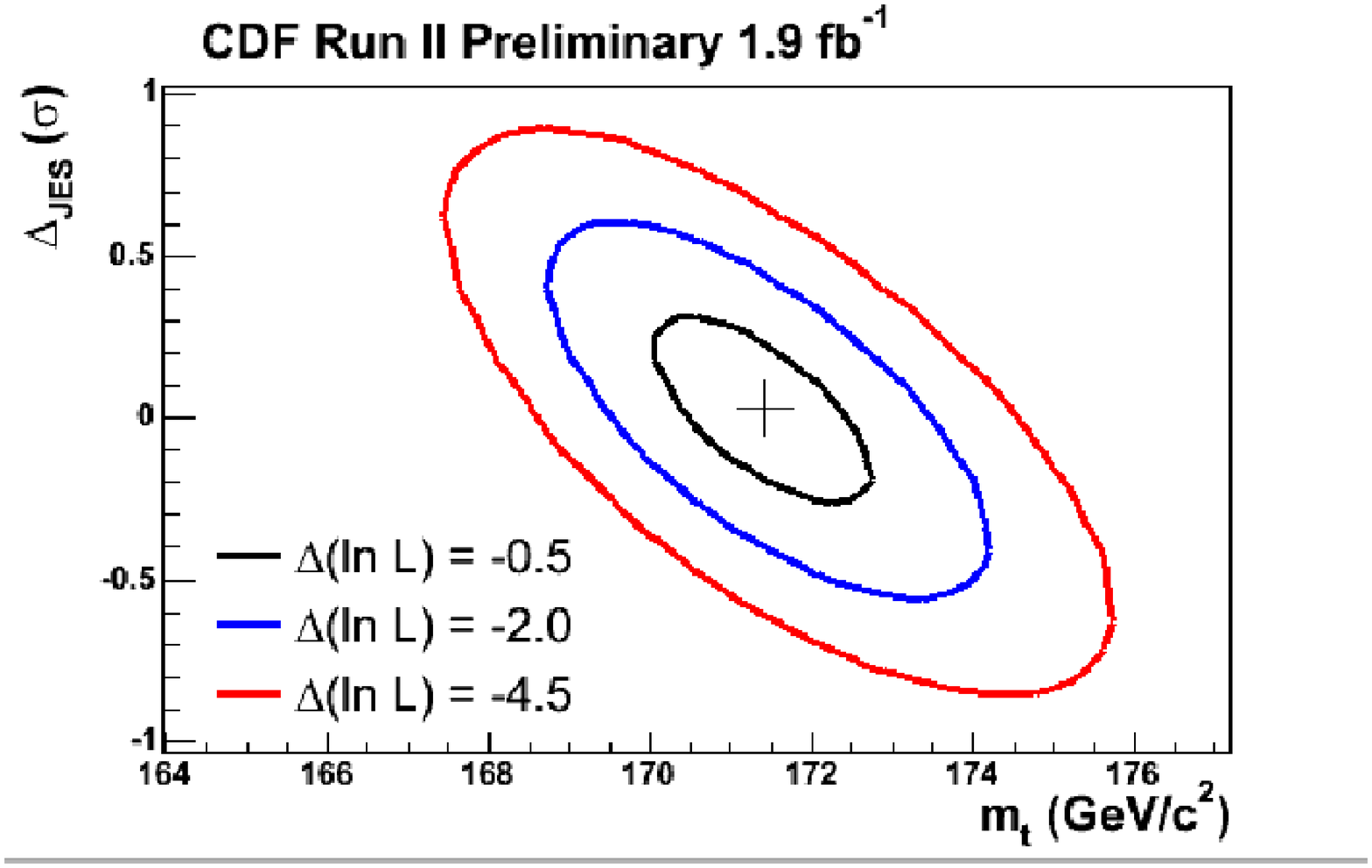}
\includegraphics[width=80.mm]{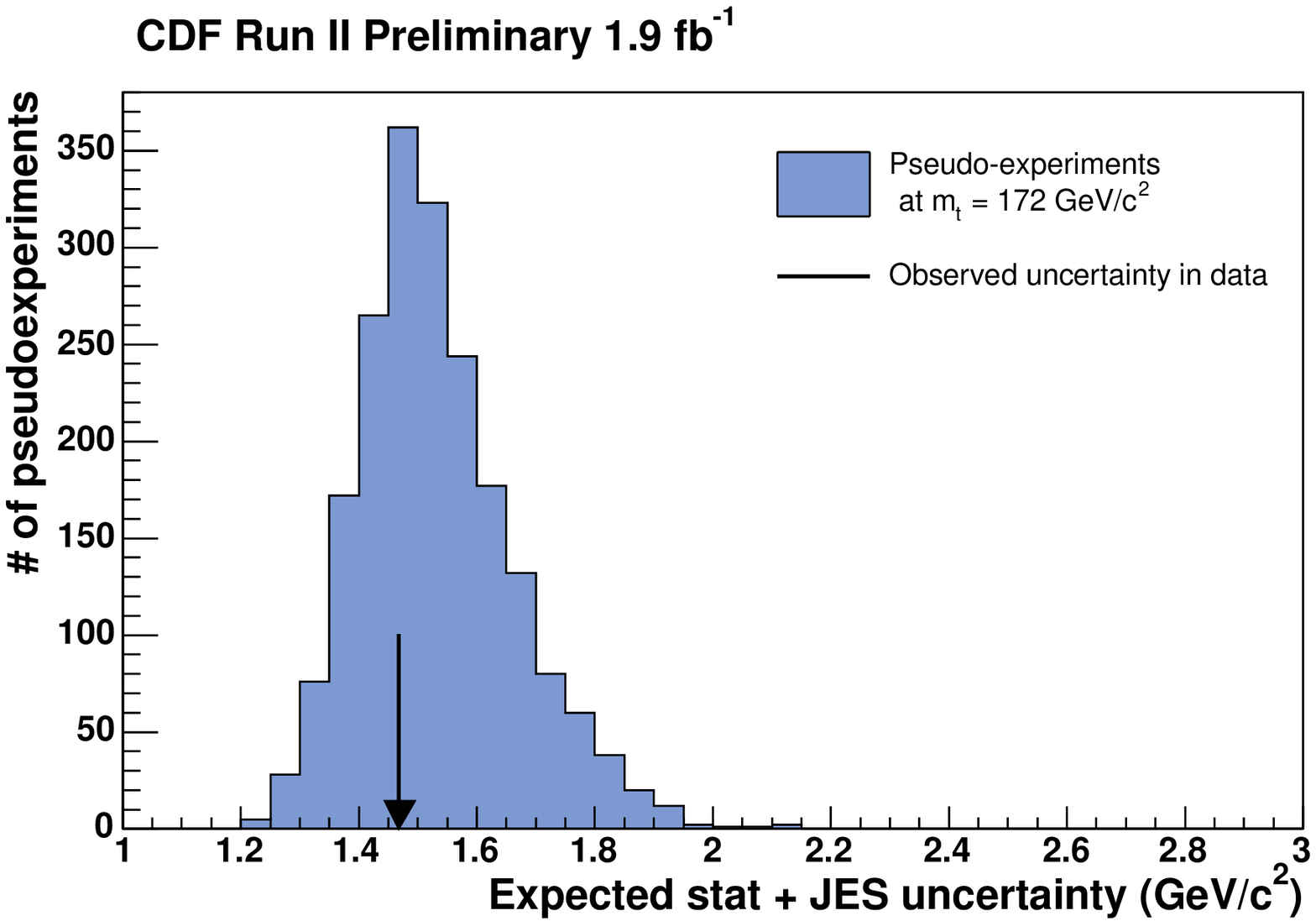}
\caption{CDF result in the \ljets channel using the ME method. 2D likelihood in
top quark mass and jet energy scale parameter for data (left).  Expected
uncertainty distribution from MC ensemble tests (right).}
\label{fig_cdf_meljets}
\end{figure*}
The second result is from CDF in the \ljets channel and is based on 1.9
fb$^{-1}$ of data~\citep{cdf_meljets}.  It also uses the ME technique and
performs an \emph{in-situ} jet energy calibration in a similar way.  Unlike the
\dzero result, it uses a neural network to discriminate between signal and
background events and, in addition, also takes the angular resolution of the
jets into account.  Figure \ref{fig_cdf_meljets} (left) shows the two
dimensional likelihood in top quark mass and jet energy scale parameter.  Figure
\ref{fig_cdf_meljets} (right) shows the expected uncertainty distribution from
MC ensemble tests with the measurement uncertainty indicated by the arrow.  The
measured result is $\mtop=171.4\pm1.5(\mbox{stat+JES})\pm1.0(\mbox{syst})$ GeV. 
The dominant sources of systematic uncertainty are the modeling of initial and
final state radiation and the residual uncertainty in the jet energy scale
representing uncertainties that cannot be addressed by a global scale factor.

\begin{figure*}[t]
\centering
\includegraphics[height=50.mm]{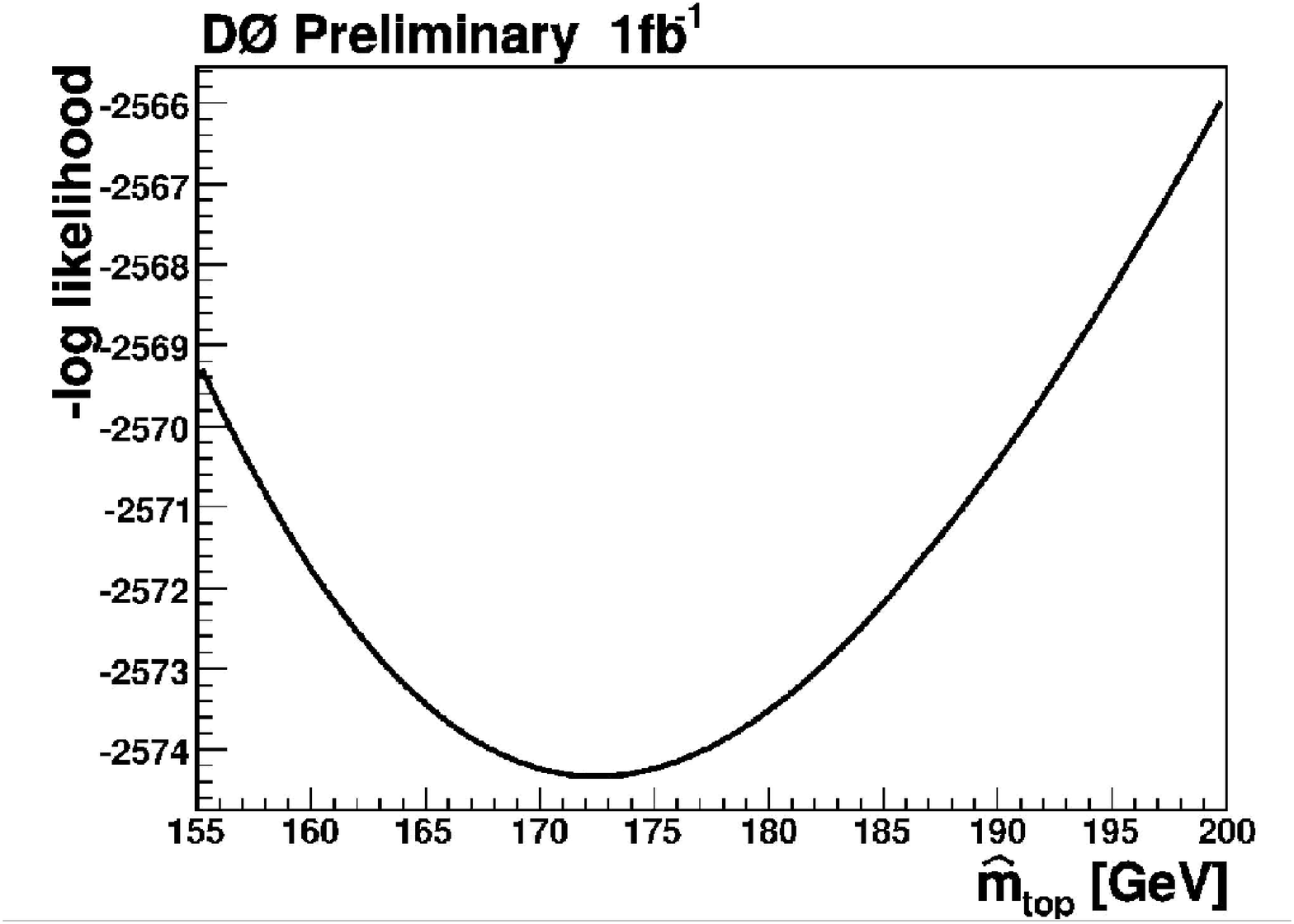}
\includegraphics[height=50.mm]{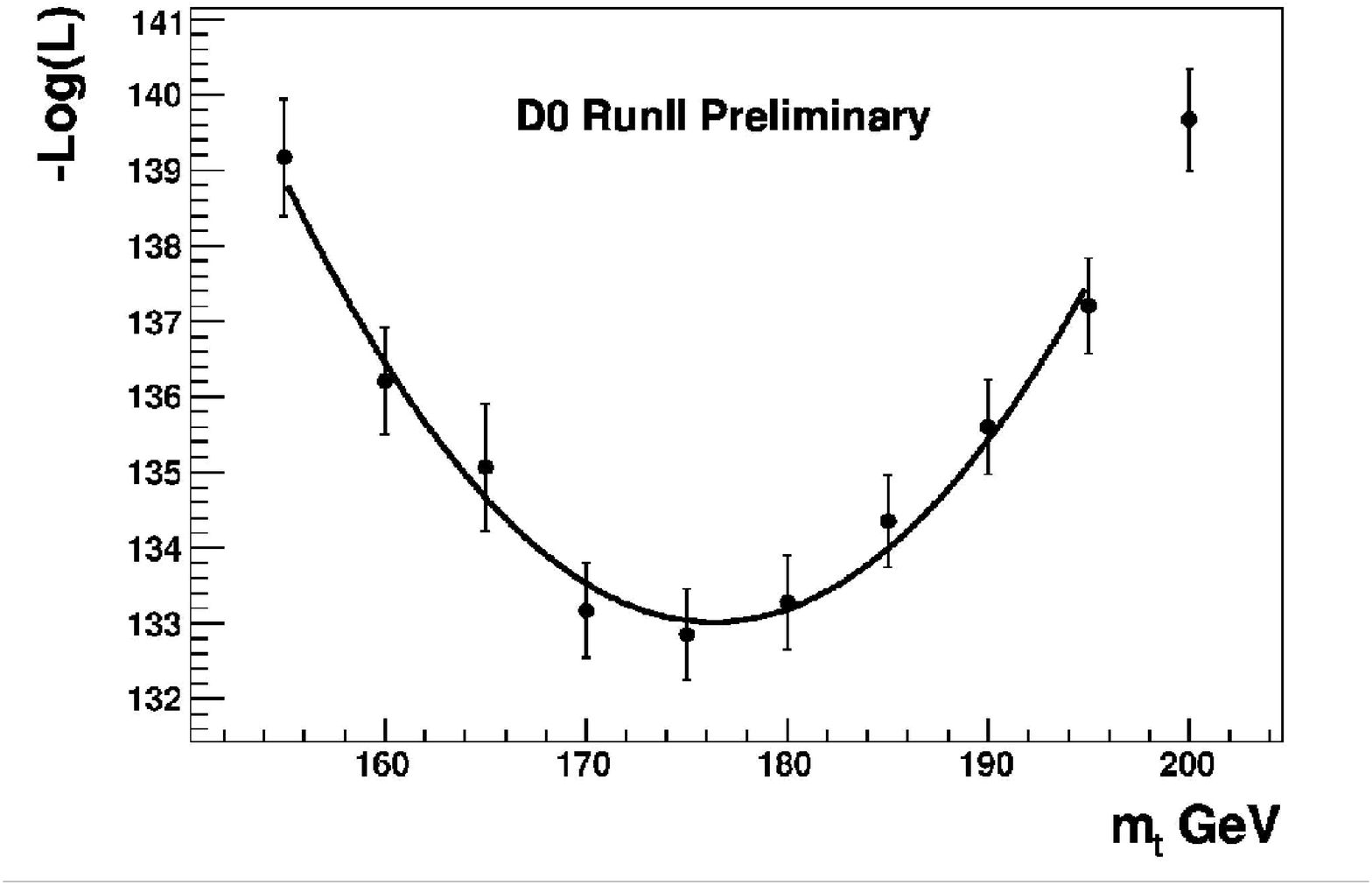}
\caption{\dzero results in the dilepton channel.  -Log likelihoods for data as a
function of top quark mass using the neutrino weighting method (left) and matrix
weighting method (right).}
\label{fig_d0_dilepton}
\end{figure*}
The third result is from \dzero in the dilepton channel and is based on 1
fb$^{-1}$ of data~\citep{d0_dilepton}.  It is a combination of two results, both
of which are based on the template method and which differ mainly in the manner
weights are assigned to the neutrino solutions.  The first is based on the
neutrino weighting technique which assigns a weight to each neutrino solution
based on the compatibility between the calculated transverse momentum of the
neutrinos and the observed missing transverse energy.  The second is based on
the matrix weighting technique which assigns weights based on the probability
for the lepton to have the observed energy in the top quark rest frame for a
given value of the top quark mass.  Negative log likelihoods as a function of
the top quark mass for both results are shown Figure \ref{fig_d0_dilepton}.
Combining both results gives a measurement of
$\mtop=173.7\pm5.4(\mbox{stat})\pm3.4(\mbox{syst})$ GeV.  The dominant sources
of uncertainty are the uncertainties in the light and $b$ jet energy scales.

\begin{figure*}[t]
\centering
\includegraphics[width=60.mm]{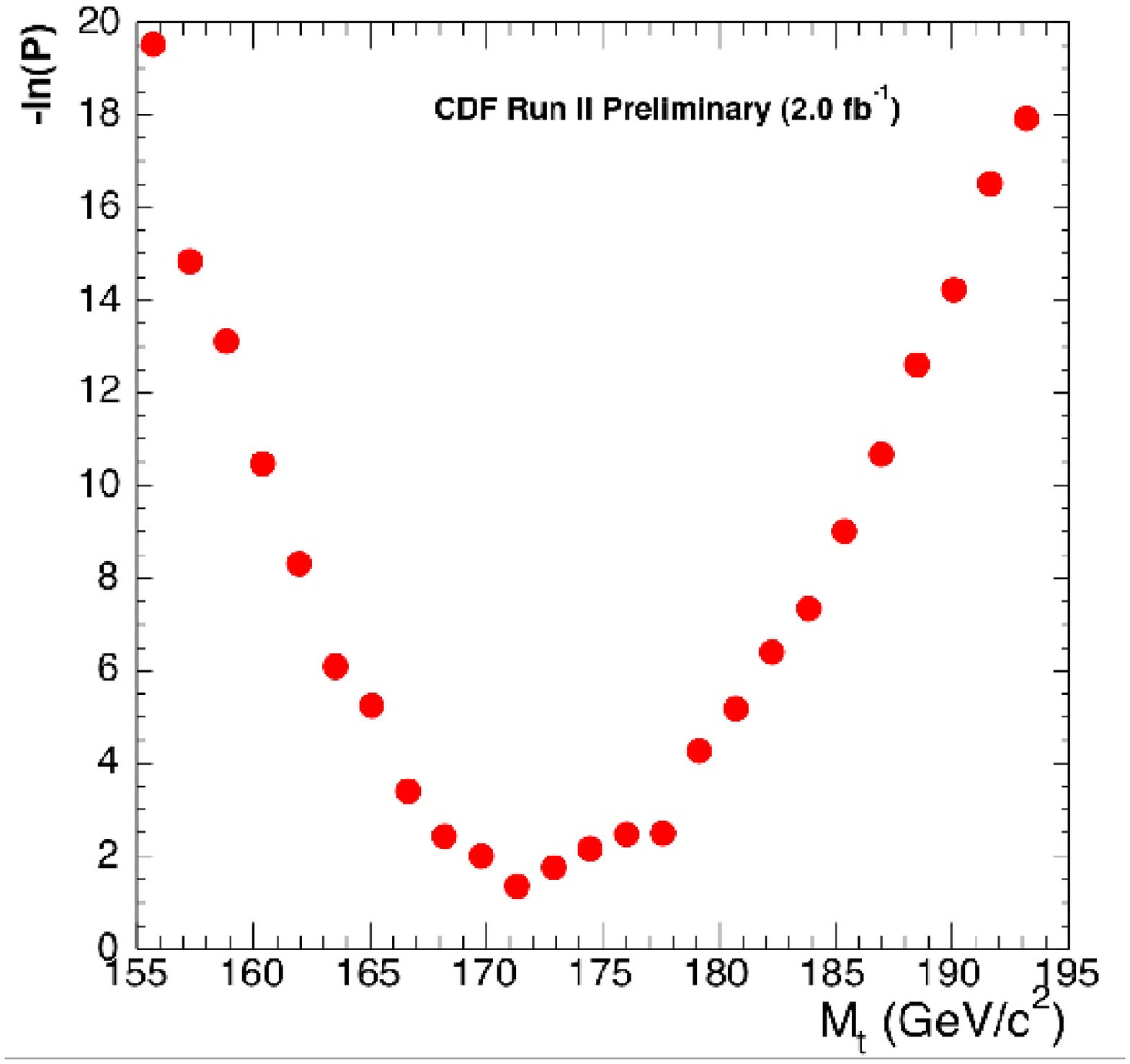}
\includegraphics[width=60.mm]{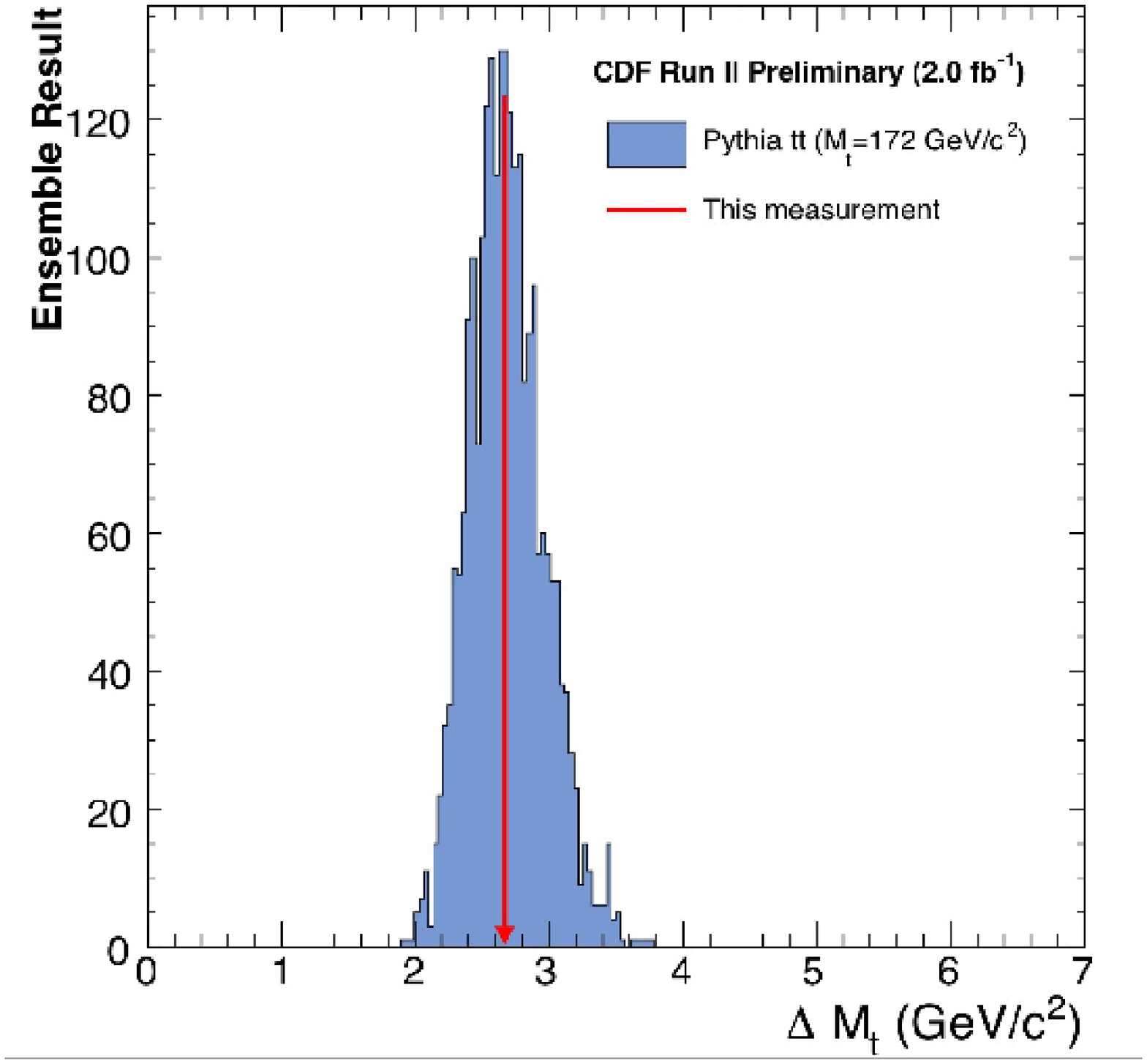}
\caption{CDF result in the dilepton channel using the ME method. -Log
likelihood for data as a function of top quark mass (left).  Distribution of
expected uncertainties from MC ensemble tests (right).}
\label{fig_cdf_dilepton}
\end{figure*}
The fourth result is also in the dilepton channel~\citep{cdf_dilepton}.  It is
from CDF and is based on 1.9 fb$^{-1}$ of data.  Like the first two lepton+jets
results, it is also based on the ME technique.  In addition, it also uses a
neural network based selection that is optimized for precision.  Figure
\ref{fig_cdf_dilepton} (left) shows the negative log likelihood as a function of
the top quark mass and Figure \ref{fig_cdf_dilepton} (right) shows the
distribution of expected uncertainties from MC ensemble tests with the arrow
indicating the measurement uncertainty.  The measured result is
$\mtop=171.2\pm2.7(\mbox{stat})\pm2.9(\mbox{syst})$ GeV.  The dominant sources
of uncertainty are the uncertainties in the jet energy scale and in the MC
generator.

\begin{figure*}[t]
\centering
\includegraphics[width=65.mm]{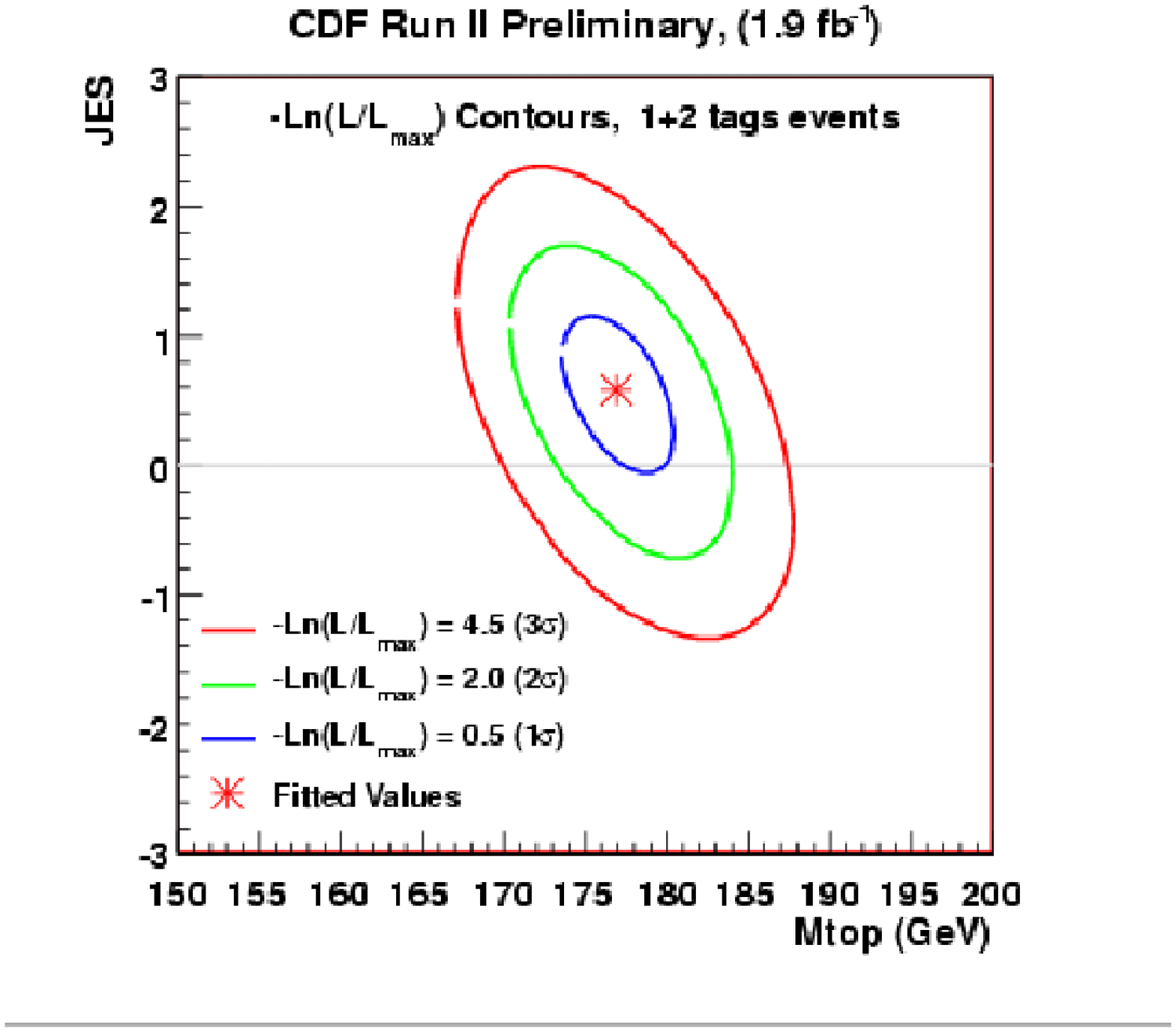}
\includegraphics[width=65.mm]{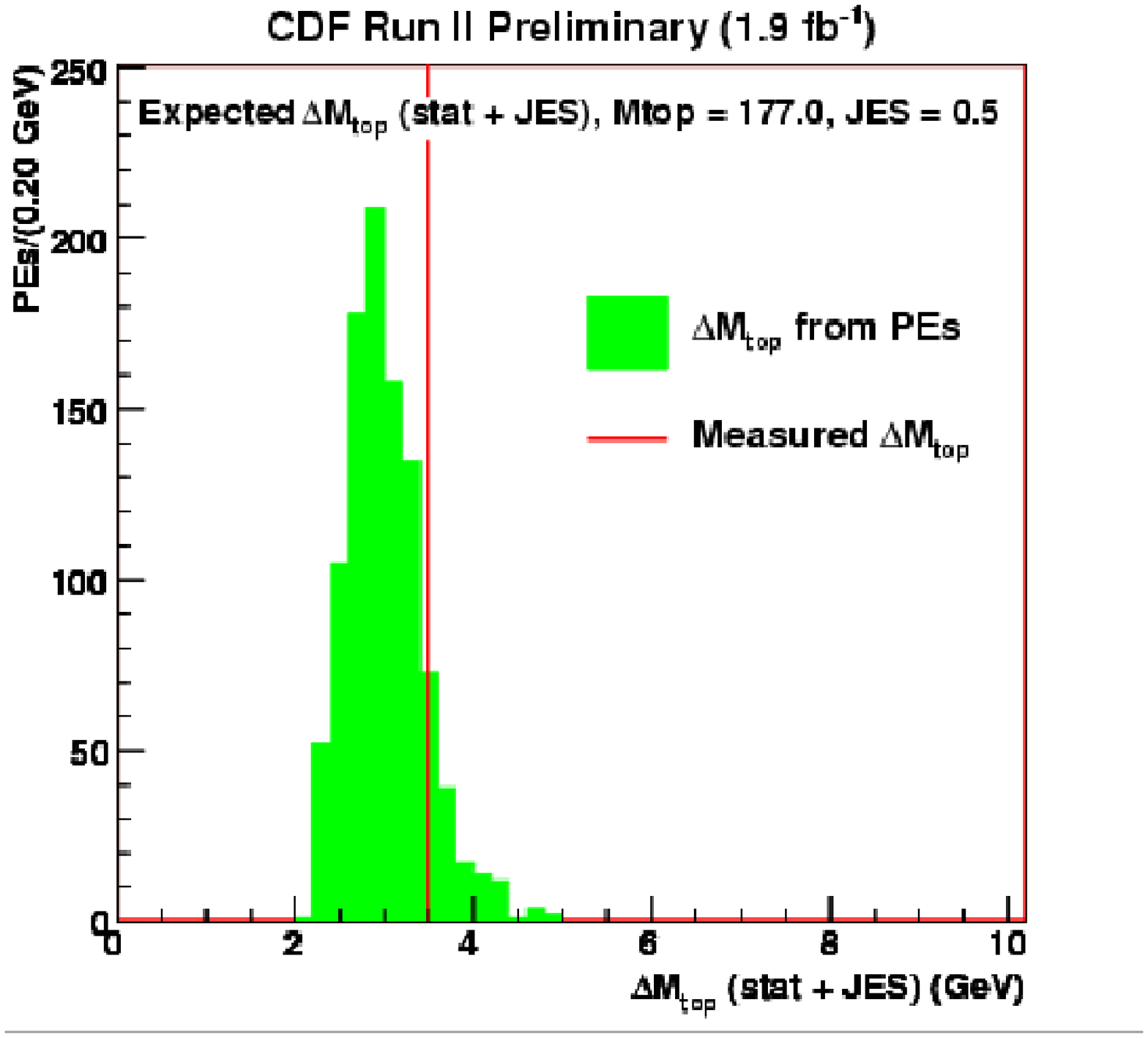}
\caption{CDF result in the all jets channel using the template method. 2D
likelihood in top quark mass and jet energy scale parameter for data (left). 
Expected uncertainty distribution from MC ensemble tests (right).}
\label{fig_cdf_alljets}
\end{figure*}
The fifth result is from CDF in the all jets channel and is based on 1.9
fb$^{-1}$ of data~\citep{cdf_alljets}.  It is based on the template method with
background shapes that are determined from data.  Like the first two \ljets
results, it also takes advantage of the $W$ mass constraint to perform an
\emph{in-situ} jet energy calibration. The two dimensional likelihood in the jet
energy scale parameter and top quark mass for data is shown in Figure
\ref{fig_cdf_alljets} (left).  The expected uncertainty distribution from MC
ensemble tests is shown in Figure \ref{fig_cdf_alljets} (right) with the
measurement uncertainty indicated by the vertical line.  The measured result is
$\mtop=177.0\pm3.7(\mbox{stat+JES})\pm1.6(\mbox{syst})$ GeV.  The systematic
uncertainty is dominated by the uncertainty in the background shapes and the
residual uncertainty in the jet energy scale not accounted for by the jet energy
scale parameter.

\begin{figure*}[t]
\centering
\includegraphics[width=67.5mm]{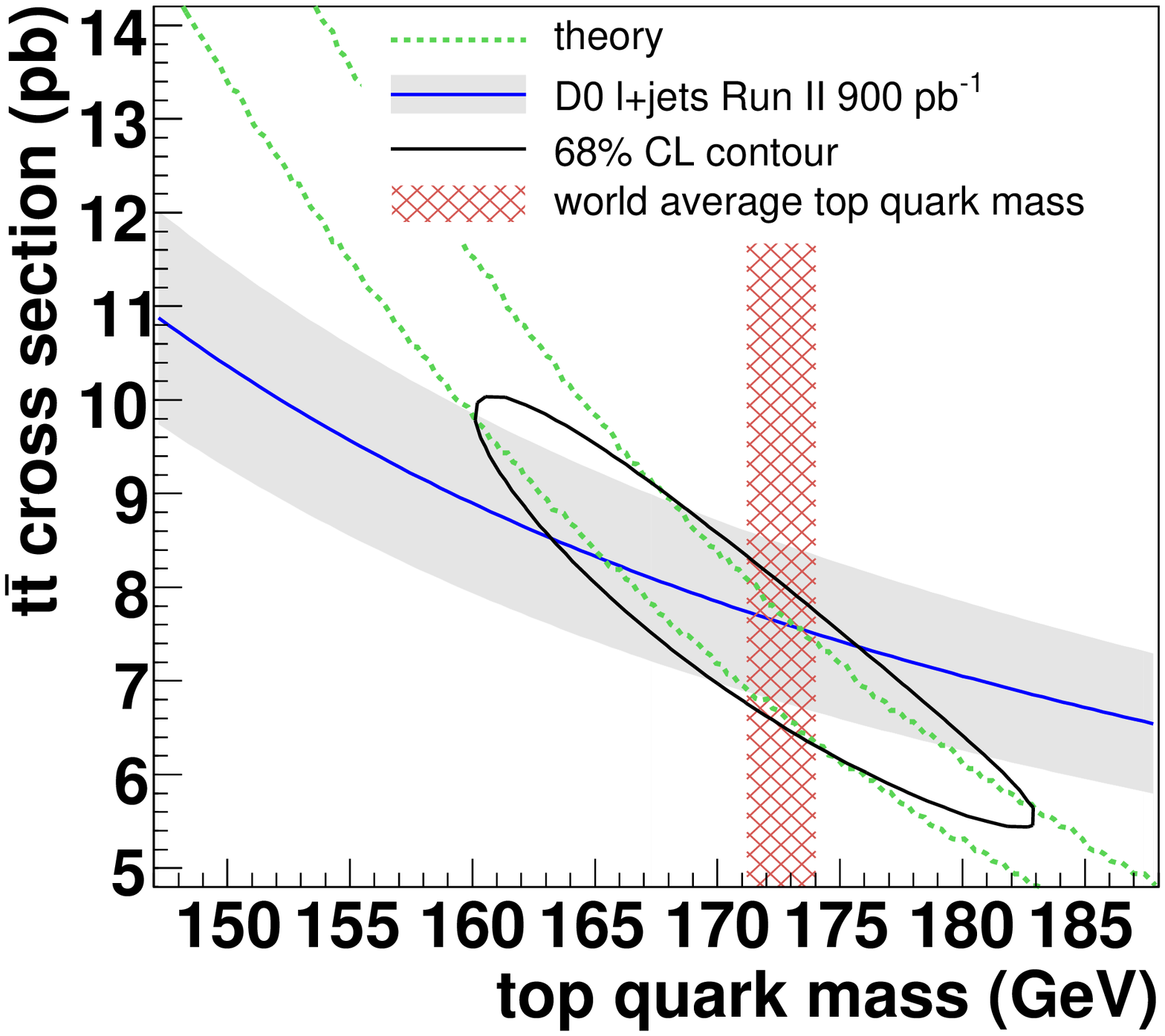}
\includegraphics[height=60.mm]{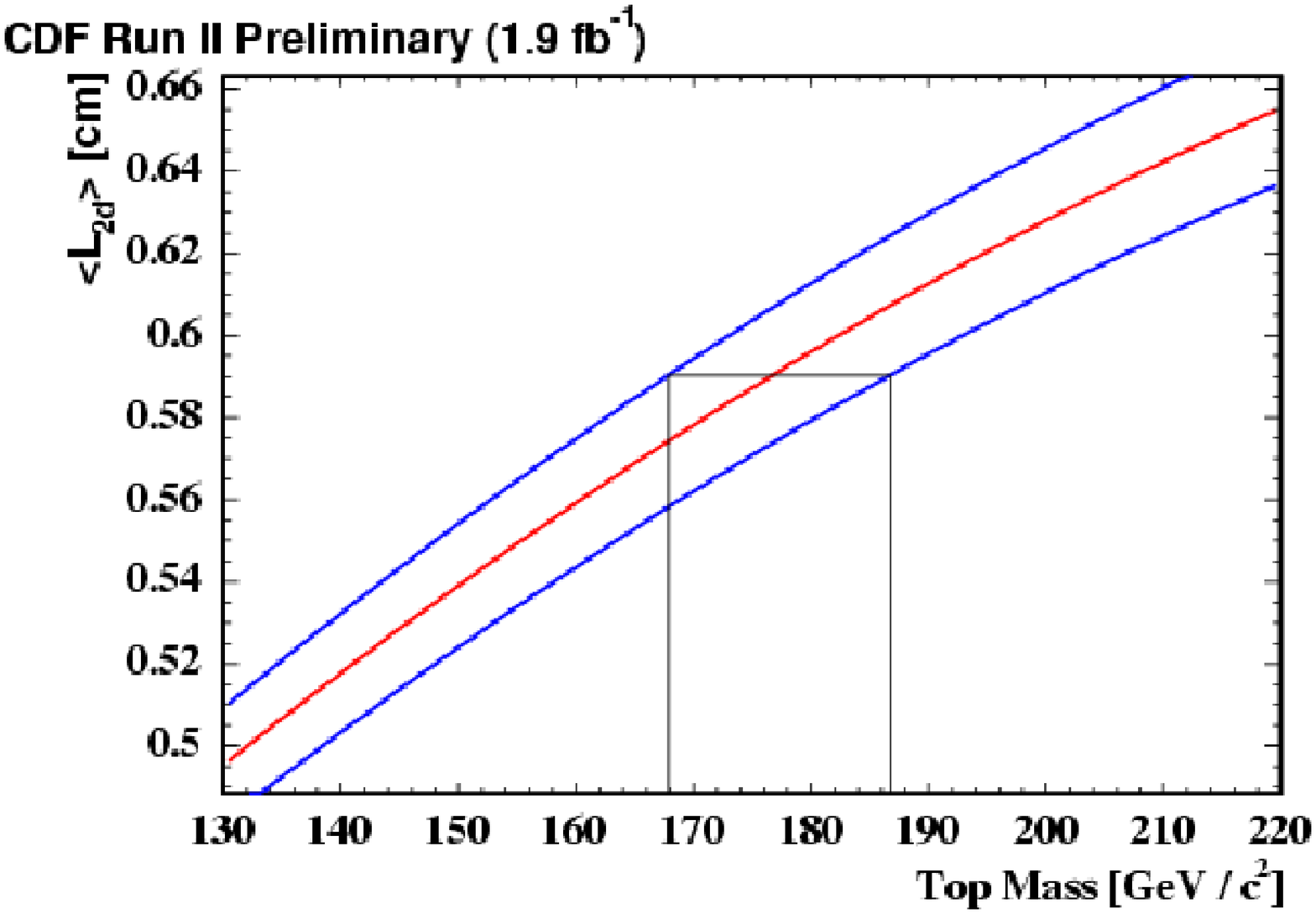}
\caption{Comparison of measured and theoretically predicted cross sections
versus top quark mass from \dzero(right). Expected average transverse decay
length including $1\sigma$ confidence intervals as a function of top quark mass
from CDF (right).}
\label{fig_novel}
\end{figure*}

The sixth result uses the indirect approach of determining the top quark mass by
comparing the theoretical and measured \ttbar production cross
sections~\citep{d0_xsect}.  It is from \dzero and is based on 1 fb$^{-1}$ of
data.  Figure \ref{fig_novel} (left) shows the theoretical and measured cross
sections as a function of the top quark mass including the 68\% C.L. contour. 
The top quark mass is found to be $\mtop=170\pm7$ GeV at 68\% C.L. in agreement
with the world average from direct measurements indicated by the cross hatched
bar in Fig. 6.

The last result uses the decay length technique to determine the top quark mass.
It is from CDF and is based on 1.9 fb$^{-1}$ of data~\citep{cdf_decaylength}. 
It is a combination of two results.  The first is based on the average
transverse decay length as described in the previous section and the second is
based on the average transverse momentum of the lepton using exactly the same
technique as that for the decay length.  Figure \ref{fig_novel} (right) shows the dependence of the
average transverse decay length on the top quark mass including one sigma
confidence intervals determined from MC events.  Also shown are the one sigma
statistical uncertainties from data represented by the horizontal bar.  The
combination of the two results gives
$\mtop=175.3\pm6.2(\mbox{stat})\pm3.0(\mbox{syst})$ GeV.
\begin{figure*}[t]
\centering
\includegraphics[width=120.mm]{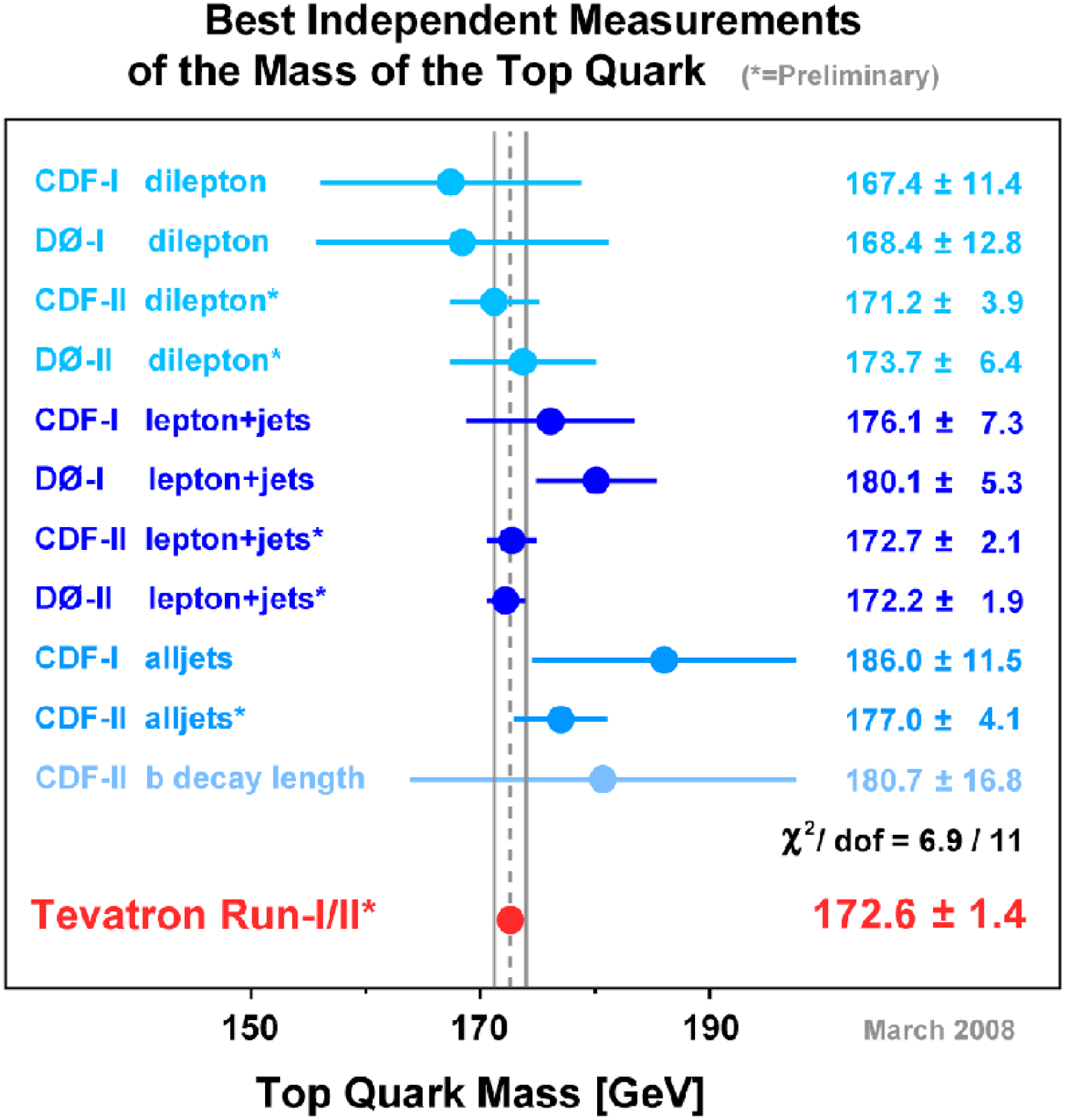}
\caption{World average top quark mass from Tevatron results.}
\label{fig_worldavg}
\end{figure*}

The world average of the top quark mass from March 2008, $\mtop=172.6\pm1.4$
GeV, based on the best Run 1 and Run 2 measurements from the Tevatron, is shown
in Fig. \ref{fig_worldavg}~\citep{worldavg}.  The uncertainty includes
statistical and systematic contributions added in quadrature.  Most of the
results presented above are included in this world average.  The exceptions are
the CDF ME \ljets (``CDF-II lepton+jets") and $b$ decay length (``CDF-II b decay
length") measurements which are from a previous version and the \dzero
measurement based on the \ttbar cross section which is not included.

\section{Conclusion}

Although the large mass of the top quark is interesting in itself, a precise
measurement of this quantity is important because of the constraint it imposes
on the Higgs mass and on possible extensions of the standard model.  Measuring
the top quark mass is a daunting task but, fortunately, various sophisticated
techniques have been developed that make the precise measurements presented here
possible.  For consistency, these measurements are performed in the different
top decay channels which are sensitive to different systematic effects and
background sources.  In addition to the ME and template based measurements, also
included are two measurements based on interesting approaches such as the \dzero
cross section technique and the CDF $b$ decay length technique.  These results
still have large uncertainties but seem promising and complement the
measurements based on the ME and template methods.  While the
results presented here are based on up to 2 fb$^{-1}$ of data, each experiment
expects to receive $\sim10$ fb$^{-1}$ by the end of 2010.  As the total uncertainty on
the world average top quark mass quickly approaches 1 GeV, it is now more
important than ever to continue refining measurement techniques.  This will
require deepening our understanding of the systematic uncertainties and
hopefully developing new ways to control the dominant ones.

\bigskip % extra skip inserted
%\begin{thebibliography}{9}   % Use for  1-9  references

\end{document}